# Size-*l* Object Summaries for Relational Keyword Search

Georgios J. Fakas*[†], Zhi Cai[†], Nikos Mamoulis[‡]

[†]Department of Computing and Mathematics
Manchester Metropolitan University, UK
{g.fakas, z.cai}@mmu.ac.uk

[‡]Department of Computer Science
The University of Hong Kong, Hong Kong
nikos@cs.hku.hk

## ABSTRACT

A previously proposed keyword search paradigm produces, as a query result, a ranked list of Object Summaries (OSs). An OS is a tree structure of related tuples that summarizes all data held in a relational database about a particular Data Subject (DS). However, some of these OSs are very large in size and therefore unfriendly to users that initially prefer synoptic information before proceeding to more comprehensive information about a particular DS. In this paper, we investigate the effective and efficient retrieval of concise and informative OSs. We argue that a good size-$l$ OS should be a stand-alone and meaningful synopsis of the most important information about the particular DS. More precisely, we define a size-$l$ OS as a partial OS composed of $l$ important tuples. We propose three algorithms for the efficient generation of size-$l$ OSs (in addition to the optimal approach which requires exponential time). Experimental evaluation on DBLP and TPC-H databases verifies the effectiveness and efficiency of our approach.

## 1. INTRODUCTION

Web Keyword Search (W-KwS) has been very successful because it allows users to extract effectively and efficiently useful information from the web using only a set of keywords. For instance, Example 1 illustrates the partial result of a W-KwS (e.g. Google) for Q1: "Faloutsos": a ranked set (with the first three results shown only) of links to web pages containing the keyword(s). We observe that each result is accompanied with a snippet [21], i.e. a short summary that sometimes even includes the complete answer to the query (if, for example, the user is only interested in whether Christos Faloutsos is a Professor or whether his brothers are academics).

The success of the W-KwS paradigm has encouraged the emergence of the keyword search paradigm in relational databases (R-KwS) [2, 4, 13]. The R-KwS paradigm is used to find tuples that contain the keywords and their relationships through foreign-key links, e.g. query Q2: "Faloutsos"+"Agrawal" returns Authors Fal-

*Partially supported by the "Hosting of Experienced Researchers from Abroad" programme (ΠΡΟΣΕΛΚΥΣΗ/ΠΡΟΕΜ/0308) funded by the Research Promotion Foundation, Cyprus.

EXAMPLE 1. *Q1 "Faloutsos" using a W-KwS (Google)*

Christos **Faloutsos**
SCS CSD Professor's affiliatons, research, projects, publications and teaching.
www.cs.cmu.edu/∼christos/ - 9k

Michalis **Faloutsos**
The Homepage of Michalis **Faloutsos** ... Interesting and Miscellaneous Links · Fun pictures · Other **Faloutsos** on the web; The Teach-To-Learn Initiative:
www.cs.ucr.edu/∼michalis/ - 5k

Petros **Faloutsos**
Courses · Press Coverage · Publications · Research Highlights · Awards · MAGIX Lab · Curriculum Vitae · Family · Other **Faloutsos** on Web.
www.cs.ucla.edu/∼pfal/ - 4k

...

EXAMPLE 2. *Q2 using an R-KwS (searching DBLP database)*
**Author**: Christos **Faloutsos**, **Paper**: Efficient similarity search in sequence databases, **Author**: Rakesh **Agrawal**.

**Author**: Christos **Faloutsos**, **Paper**: Method for high-dimensionality indexing in a multi-media database, **Author**: Rakesh **Agrawal**.

**Author**: Christos **Faloutsos**, **Paper**: Quest: A project on database mining, **Author**: Rakesh **Agrawal**.

EXAMPLE 3. *Q1 using an R-KwS (searching DBLP database)*
**Author**: Christos **Faloutsos**
**Author**: Michalis **Faloutsos**
**Author**: Petros **Faloutsos**

outsos and Agrawal and their associations through co-authored papers. Example 2 illustrates the result of a traditional R-KwS for Q2 on the DBLP database. On the other hand, the R-KwS paradigm may not be very effective when trying to extract information about a particular *data subject* (DS), e.g. "Faloutsos" in Q1. Example 3 illustrates the R-KwS result for Q1, namely a ranked set of Author tuples containing the Faloutsos keyword, which are the Author tuples corresponding to the three brothers. Evidently, this result fails to provide comprehensive information to users about the Faloutsos brothers, e.g. a complete list of their publications and other corresponding details (Certainly, the R-KwS paradigm remains very useful when trying to combine keywords).

In [8], the concept of *object summary* (OS) is introduced; an OS summarizes all data held in a database about a particular DS. More precisely, an OS is a tree with the tuple $t^{DS}$ containing the keyword (e.g. Author tuple Christos Faloutsos) as the root node and its neighboring tuples, containing additional information (e.g. his papers, co-authors etc.), as child nodes. The result for Q1 is in fact a set of OSs: one per DS that includes all data held in the database for each Faloutsos brother. Example 4 illustrates the OS for Christos (the complete set of papers and the OSs of the other two brothers were omitted due to lack of space). This result evidently provides a more complete set of information per brother.

EXAMPLE 4. *The OS for Christos Faloutsos*

**Author**: Christos **Faloutsos**
  **Paper**: On Power-law Relationalships of the Internet Topology.
    **Conference**:SIGCOMM. **Year**:1999.
    **Co-Author(s)**:Michalis Faloutsos, Petros Faloutsos.
  **Paper**: An Efficient Pictorial Database System for PSQL.
    **Conference**:IEEE Trans. Software Eng. **Year**:1988.
    **Co-Author(s)**:N. Roussopoulos, T. Sellis.
  ...
  ...
  **Paper**: Declustering Using Fractals.
    **Conference**:PDIS. **Year**:1993. **Co-Author(s)**:Pravin Bhagwat.
  **Paper**: Declustering Using Error Correcting Codes.
    **Conference**:PODS. **Year**:1989. **Co-Author(s)**:Dimitris N. Metaxas.
**(Total 1,309 tuples)**

EXAMPLE 5. *The size-l OSs for Q1 and l=15*

**Author**: Christos **Faloutsos**
  **Paper**: On Power-law Relationalships of the Internet Topology.
    **Conference**:SIGCOMM. **Year**:1999.
    **Co-Author(s)**:Michalis Faloutsos, Petros Faloutsos.
  **Paper**: The QBIC Project: Querying Images by Content, Using, Color, Texture and Shape.
    **Conference**:SPIE. **Year**:1993.
    **Co-Author(s)**:Carlton W. Niblack, Dragutin Petkovic, Peter Yanker.
  **Paper**: Efficient and Effective Querying by Image Content.
    **Conference**:J. Intell. Inf. Syst. **Year**:1994.
    **Co-Author(s)**:N. Roussopoulos, T. Sellis.
  ...
**Author**: Michalis **Faloutsos**
  **Paper**: On Power-law Relationalships of the Internet Topology.
    **Conference**:SIGCOMM. **Year**:1999.
    **Co-Author(s)**:Christos Faloutsos, Petros Faloutsos.
  **Paper**: QoSMIC: Quality of Service Sensitive Multicast Internet Protocol.
    **Conference**:SIGCOMM. **Year**:1998.
    **Co-Author(s)**:Anindo Banerjea, Rajesh Pankaj.
  **Paper**: Aggregated Multicast with Inter-Group Tree Sharing.
    **Conference**:Networked Group Communication. **Year**:2001.
    **Co-Author(s)**:Aiguo Fei.
  ...
**Author**: Petros **Faloutsos**
  **Paper**: On Power-law Relationalships of the Internet Topology.
    **Conference**:SIGCOMM. **Year**:1999.
    **Co-Author(s)**:Christos Faloutsos, Michalis Faloutsos.
  **Paper**: Composable controllers for physics-based character animation.
    **Conference**:SIGGRAPH. **Year**:2001.
    **Co-Author(s)**:Michiel van de Panne, Demetri Terzopoulos.
  **Paper**: The virtual stuntman: dynamic characters with a repertoire of autonomous motor skills.
    **Conference**:Computers & Graphics 25. **Year**:2001.
    **Co-Author(s)**:Michiel van de Panne, Demetri Terzopoulos.

From Example 4, we can observe that some of the OSs may be very large in size; e.g. Christos Faloutsos has co-authored many papers and his OS consists of 1,309 tuples. This is not only unfriendly to users that prefer a quick glance first before deciding which Faloutsos they are really interested in, but also expensive to produce. Therefore, a *partial* OS of size $l$, composed of only $l$ representative and important tuples, may be more appropriate.

In this paper, we investigate in detail the effective and efficient generation of size-$l$ OSs. Example 5 illustrates Q1 with $l$=15 on the DBLP database; namely a set of size-15 OSs composed of only 15 important tuples for each DS. From the user's perspective, the semantics of this paradigm resemble more a W-KwS rather than a R-KwS. For instance, the complete OS of Example 4 resembles a web page (as they both include comprehensive information about a DS), whereas the size-$l$ OSs of Example 5 resemble the snippets of Example 1. Therefore, users with W-KwS experience will potentially find it friendlier and also closer to their expectations.

OSs and size-$l$ OSs can have many applications. For example, OSs can automate responds to data protection act (DPA) subject access requests (e.g. the US Privacy Act of 1974, UK DPA of 1984 and 1998 [1] etc.). According to DPA access requests, DSs have the right to request access from any organization to personal information about them. Thus, data controllers of organizations must extract data for a given DS from their databases and present it in an intelligible form [10]. Another application is for intelligent services searching information about suspects from various databases. Hence, size-$l$ OSs can also be very useful as they enhance the usability of OSs. In general, a size-$l$ OS is a concise summary of the *context* around any pivot database tuple, finding application in (interactive) data exploration, schema extraction, etc.

We should effectively generate a stand-alone size-$l$ OS, composed of $l$ important tuples only, so that the user can comprehend it without any additional information. A stand-alone size-$l$ OS should preserve meaningful and self-descriptive semantics about the DS. As we explain in Section 3, for this reason, the $l$ tuples should form a connected graph that includes the root of the OS (i.e. $t^{DS}$). To distinguish the importance of individual tuples $t_i$ to be included in the size-$l$ OS, a *local importance* score (denoted as $Im(OS, t_i)$) is defined by combining the tuple's *global importance* score in the database (denoted as $Im(t_i)$) and its *affinity* [8] in the OS (denoted as $Af(t_i)$). Based on the local importance scores of the tuples of an OS, we can find the partial OS of size $l$ with the maximum importance score, which includes tuples that are connected with $t^{DS}$.

The efficient size-$l$ generation of OSs is a challenging problem. A brute force approach, that considers all candidate size-$l$ OSs before finding the one with the maximum importance, requires exponential time. We propose an optimal algorithm based on dynamic programming, which is efficient for small problems, however, it does not scale well with the OS size and $l$. In view of this, we design three practical greedy algorithms.

We provide an extensive experimental study on DBLP and TPC-H databases, which includes comparisons of our algorithms and verifies their efficiency. To verify the effectiveness of our framework, we collected user feedback, e.g. by asking several DBLP authors (i.e. the DSs themselves) to assess the computed size-$l$ OSs of themselves on the DBLP database. The users suggested that the results produced by our method are very close to their expectations.

The rest of the paper is structured as follows. Section 2 describes background and related work. Section 3 describes the semantics of size-$l$ OS keyword queries and formulates the problem of their generation. Sections 4 and 5 introduce the optimal and greedy algorithms respectively. Section 6 presents experimental results and Section 7 provides concluding remarks.

## 2. BACKGROUND AND RELATED WORK

In this section, we first describe the concept of *object summaries* (OSs), which we build upon in this paper. We then present and compare other related work in R-KwS, ranking and summarization. To the best of our knowledge there is no previous work that focuses on the computation of size-$l$ OSs.

### 2.1 Object Summaries

In the context of OS search in relational databases [8, 7], a query is a set of keywords (e.g. "Christos Faloutsos") and the result is a set of OSs. An OS is generated for each tuple ($t^{DS}$) found in the database that contains the keyword(s) as part of an attribute's value (e.g. tuple "Christos Faloutsos" of relation Author in the DBLP database). An OS is a tree structure composed of tuples, having $t^{DS}$ as root and $t^{DS}$'s neighboring tuples (i.e. those associated through foreign keys) as its children/descendants.

In order to construct OSs, this approach combines the use of graphs and SQL. The rationale is that there are relations, denoted as $R^{DS}$ (e.g. the Author relation), which hold information about the queried Data Subjects (DSs) and the relations linked around $R^{DS}$s contain additional information about the particular DS. For each $R^{DS}$, a Data Subject Schema Graph ($G^{DS}$) can be generated; this is a directed labeled tree that captures a subset of the database schema with $R^{DS}$ as a root. (Figures 1 and 11 illustrate the schemata of the DBLP and TPC-H databases whereas Figures 2 and 12 illustrate exemplary $G^{DS}$s.) Each relation in $G^{DS}$ is also annotated with useful information that we describe later, such as affinity and importance. $G^{DS}$ is a "treealization" of the schema, i.e. $R^{DS}$ becomes the root, its neighboring relations become child nodes and also any looped or many-to-many relationships are replicated. Examples of such replications are relations PaperCitedBy, PaperCites and Co-Author on Author $G^{DS}$ and relations Partsupp, Lineitem, Parts etc. on Customer $G^{DS}$ (see $G^{DS}$s in Figures 2 and 12). (User evaluation in [8] verified that the tree format (achieved via such replications) increases significantly friendliness and ease of use of OSs.) The challenge now is the selection of the relations from $G^{DS}$ which have the highest affinity with $R^{DS}$; these need to be accessed and joined in order to create a good OS. To facilitate this task, affinity measures of relations (denoted as $Af(R_i)$) in $G^{DS}$ are investigated, quantified and annotated on the $G^{DS}$. The affinity of a relation $R_i$ to $R^{DS}$ can be calculated using the following formula:

$$Af(R_i) = \sum_j m_j w_j \cdot Af(R_{Parent}), \quad (1)$$

where $j$ ranges over a set of metrics $(m_1, m_2, \ldots, m_n)$, their corresponding weights $(w_1, w_2, \ldots, w_n)$ and $Af(R_{Parent})$ is the affinity of $R_i$'s *parent* to $R^{DS}$. Affinity metrics between $R_i$ and $R^{DS}$ include (1) their distance and (2) their connectivity properties on both the database schema and the data-graph (see [8] for more details). Given an affinity threshold $\theta$, a subset of $G^{DS}$ can be produced, denoted as $G^{DS}(\theta)$. Finally, by traversing $G^{DS}(\theta)$ (e.g. by joining the corresponding relations) we can generate the OSs (either by using the precomputed data-graph or directly from the database using Algorithm 5). More precisely, a breadth-first traversal of the corresponding $G^{DS}(\theta)$ with the $t^{DS}$ tuple as the initial root entry of the OS tree is applied. For instance, for keyword query Q1, Author $G^{DS}$ of Figure 2 and $\theta=0.7$ the report presented in Example 4 will automatically be generated. Note that Author $G^{DS}(0.7)$ includes all relations whilst Customer $G^{DS}(0.7)$ includes only Customer, Nation, Region, Order, Lineitem and Partsupp relations (since all these relations have affinity greater than 0.7). Similarly, the set of attributes $A_j$ from each relation $R_i$ that are included in a $G^{DS}$ are selected by employing an *attributes affinity* and a threshold (i.e. $\theta'$). For example, in a Customer OS, Comment is excluded from Partsupp relation as it is not relevant to Customer DSs.

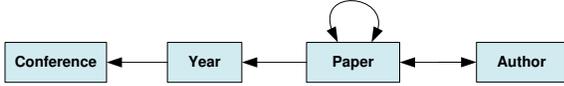

**Figure 1: The DBLP Database Schema**

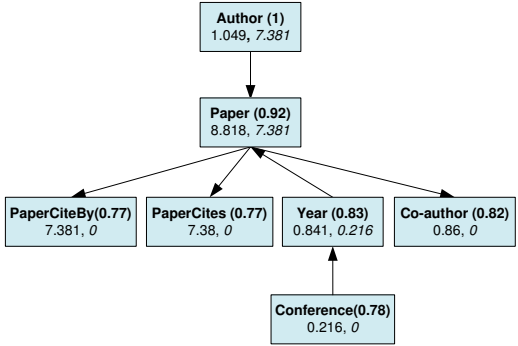

**Figure 2: The DBLP Author $G^{DS}$ (Annotated with (Affinity), max($R_i$) and mmax($R_i$))**

## 2.2 R-KwS and Ranking

R-KwS techniques facilitate the discovery of joining tuples (i.e. Minimal Total Join Networks of Tuples (MTJNTs) [13]) that collectively contain all query keywords and are associated through their keys; for this purpose the concept of *candidate networks* is introduced; see, for example, DISCOVER [13], BANKS [2, 4]. The OSs paradigm differs from other R-KwS techniques semantically, since it does not focus on finding and ranking candidate networks that connect the given keywords, but searches for OSs, which are trees centered around the data subject described by the keywords.

Précis Queries [15, 19] resemble size-$l$ OSs as they append additional information to the nodes containing the keywords, by considering neighboring relations that are implicitly related to the keywords. More precisely, a précis query result is a logical subset of the original database (i.e. a subset of relations and a subset of tuples). For instance, the précis of Q1 is a subset of the database that includes the tuples of the three Faloutsos Authors and a subset of their (common) Papers, Co-Authors, Conferences, etc. In contrast, our result is a set of three separate size-$l$ OSs (Example 5). A thorough evaluation between OSs and précis appears in our earlier work [8].

R-KwS techniques also investigate the ranking of their results. Such ranking paradigms consider:

**1) IR-Style techniques**, which weight the amount of times keywords (terms) appear in MTJNs [12, 16, 17, 23]. However, such techniques miss tuples that are related to the keywords, but they do not contain them [3]; e.g. for Q1, tuples in relation Papers also have importance although they do not include the Faloutsos keyword.

**2) Tuples' Importance**, which weights the authority flow through relationships, e.g. ObjectRank [3], [22], ValueRank [9], PageRank [5], BANKS (PageRank inspired) [2], [4], XRANK [11] etc. In this paper we use tuples' importance to model global importance scores and more precisely global ObjectRank (for DBLP) and ValueRank (for TPC-H). (Note that our algorithms are orthogonal to how tuple importance is defined and other methods could also be investigated.) ObjectRank [3] is an extension of PageRank on databases and introduces the concept of Authority Transfer Rates between the tuples of each relation of the database (Authority Transfer Rates are annotated on the so called Authority Transfer Schema Graph, denoted as $G^A$, e.g. Figure 13). They are based on the observation that solely mapping a relational database to a graph (as in the case of the web) is not accurate and a $G^A$ is required to control the flow of authority in neighboring tuples. For instance, well cited papers should have higher importance than papers citing many other papers or a well cited paper should have better ranking than another one with fewer citations. ValueRank is an extension of ObjectRank which also considers the tuples' values and thus can be applied on any database (e.g. TPC-H) in contrast to ObjectRank which is mainly effective on authoritative flow data such as bibliographic data (e.g. DBLP). For instance, in trading databases, a customer with five orders of values $10 may get lower importance than another customer with three orders of values $100.

## 2.3 Other Related Work

Document summarization techniques have attracted significant research interest [20, 21]. In general, these techniques are IR-style inspired. Web snippets [21] are examples of document summaries that accompany search results of W-KwSs in order to facilitate their quick preview (e.g. see Example 1). They can be either static (e.g.

composed of the first words of the document or description metadata) or query-biased (e.g. composed of sentences containing many times the keywords) [20]. Still, the direct application of such techniques on databases in general and OS in particular is ineffective; e.g. they disregard the relational associations and semantics of the displayed tuples. For example consider Q1 and Example 4, papers authored by Faloutsos (although don't include the Faloutsos keyword) have importance analogous to their citations and authors; this is ignored by document summarization techniques.

XML keyword search techniques, similarly to R-KwSs, facilitate the discovery of XML sub-trees that contain all query keywords (e.g. "Faloutsos"+"Agrawal"). Analogously, XML snippets [14] are sub-trees of the complete XML result, with a given size, that contain all keywords. An apparent difference between size-$l$ OSs and XML snippets is their semantics which is analogous to the semantic difference between complete OSs and XML results. Therefore, their generation is a completely different problem. An interesting similarity is that both size-$l$ OS and XML snippets are sub-trees of the corresponding complete results, hence composed of connected nodes. This common property is for the same reason, i.e. to preserve self-descriptiveness.

## 3. SIZE-*L* OS

A size-$l$ OS keyword query consists of (1) a set of keywords and (2) a value for $l$ (e.g. Q1 and $l$=15) and the result comprises a set of size-$l$ OSs. A good size-$l$ OS should be a stand-alone and meaningful synopsis of the most important information about the particular DS.

DEFINITION 1. *Given an OS and an integer size l, a candidate size-l OS is any subset of the OS composed of l tuples, such that all l tuples are connected with $t^{DS}$ (i.e. the root of the OS tree).*

Definition 1 guarantees that the size-$l$ OS remains stand-alone, (so users can understand it as it is without any additional tuples); i.e. by including connecting tuples we also include the semantics of their connection to the DS. (Recall that this criterion was also used in [14] for the same reasons.) Consider the example of Figure 3 which is a fraction of the Faloutsos OS (in the DBLP database). Even, if the Paper "Efficient and Effective Querying by Image Content" has less local importance (e.g. 20) than the Co-Author(s) Sellis (e.g. 43) and Roussopoulos (e.g. 34), we cannot exclude the Paper and include only the Co-Authors. The rationale is that by excluding the Paper tuple we also exclude the semantic association between the Author and Co-Author(s), which in this case is their common paper. Also note that a size-$l$ OS will not necessarily include the $l$ tuples with the largest importance scores. For example, the Co-Author Roussopoulos, although with larger importance than the particular Paper, may have to be excluded from a size-$l$ OS (e.g. from a size-3 OS which will consist of (1) Author "Faloutsos", (2) Paper "Efficient . . . " and (3) Co-Author "Sellis").

Given an OS, we can extract exponentially many size-$l$ OSs that satisfy Definition 1. In the next section we define a measure for the *importance* (i.e., quality) of a candidate size-$l$ OS. Our goal then would be to retrieve a size-$l$ OS of the maximum possible quality.

### 3.1 Importance of a Size-*l* OS

The (global) importance of any candidate size-$l$ OS $S$, denoted as $Im(S)$, is defined as:

$$Im(S) = \sum_{t_i \in S} Im(OS, t_i), \quad (2)$$

where $Im(OS, t_i)$ is the *local importance* of tuple $t_i$ (to be defined in Section 3.2 below). We say that a candidate size-$l$ OS

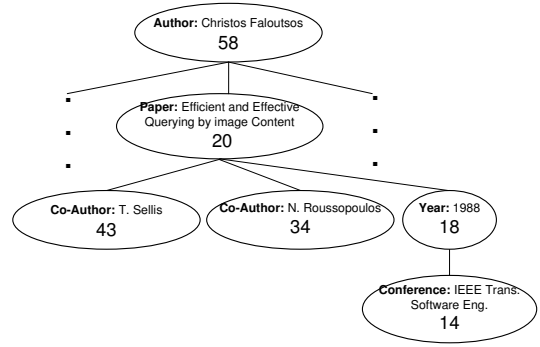

**Figure 3: A Fraction of the Faloutsos OS (Annotated with Local Importance)**

is an *optimal* size-$l$ OS, if it has the maximum $Im(S)$ (denoted as $max(Im(S))$) over all candidate size-$l$ OSs for the given OS. Wherever an optimal size-$l$ OS is hard to find, we target the retrieval of a sub-optimal size-$l$ OS of the highest possible importance.

### 3.2 Local Importance of a Tuple *($Im(OS, t_i)$)*

The local importance of $Im(OS, t_i)$ of each tuple $t_i$ in an OS can be calculated by:

$$Im(OS, t_i) = Im(t_i) \cdot Af(t_i), \quad (3)$$

where $Im(t_i)$ is the global importance of $t_i$ in the database. We use global ObjectRank and ValueRank to calculate global importance, as discussed in Section 2.2. $Af(t_i)$ is the affinity of $t_i$ to the $t^{DS}$; namely the affinity $Af(R_i)$ of the corresponding relation $R_i$ where $t_i$ belongs, to $R^{DS}$. This can be calculated from $G^{DS}$ using Equation 1, as discussed in Section 2.1 (alternatively, a domain expert can set $Af(R_i)$s manually). For example, if tuple $t_i$ is paper "Efficient.." with $Im(t_i)$=21.74 and $Af(t_i)$=$Af(R_{Paper})$=0.92 (see the affinity on Author $G^{DS}$ in Figure 2), then $Im(OS, t_i)$= 21.74*0.92=20.

Multiplying global importance $Im(t_i)$ with affinity $Af(t_i)$ reduces the importance of tuples that are not closely related to the DS. For instance, although paper "Efficient .." and year "1988" have equal global importance scores (21.74 and 21.64, respectively), their local importance scores become 20 (=21.74*0.92) and 18 (=21.64*0.83) respectively. The use of importance and affinity metrics is inspired by other earlier work; e.g. XRANK and précis employ variations of importance and affinity [11, 15]. For defining affinity in [11, 15], only distance is considered; however, as it is shown in [8] distance is only one among the possible affinity metrics (e.g. cardinality, reverse cardinality etc.).

### 3.3 Problem Definition

The generation of a complete OS is straightforward: we only have to traverse the corresponding $G^{DS}$ (see Algorithm 5 in the Appendix). The generation of a size-$l$ OS is a more challenging task because we need to select $l$ tuples that are connected to the $t^{DS}$ of the tree and at the same time result to the largest $Im(S)$. Hence, the problem we study in this paper can be defined as follows:

PROBLEM 1 (FIND AN OPTIMAL SIZE-$l$ OS). *Given a $t^{DS}$, the corresponding $G^{DS}$ and l, find a size-l OS S of maximum $Im(S)$.*

A direct approach for solving this problem is to first generate the complete OS (i.e. Algorithm 5)[1] and then determine the optimal

---
[1]In fact, any tuples or subtrees, which have distance at least $l$ from the root $t^{DS}$ are excluded from the OS, as these cannot be part of a connected size-$l$ OS rooted at $t^{DS}$.

size-$l$ OS from it. In Section 4, we propose a dynamic programming (DP) algorithm for this purpose. If the complete OS is too large, solving the problem exactly using DP can be too expensive. In view of this, in Section 5, we propose greedy algorithms that find a sub-optimal synopsis. In order to further reduce the cost of finding a sub-optimal solution, in Section 5.3, we also propose an economical approach, which, instead of the complete OS, initially generates a preliminary partial OS, denoted as prelim-$l$ OS. The rationale of a prelim-$l$ OS is to avoid the extraction and consequently further processing of fruitless tuples that are not promising to make it in the size-$l$ OS. DP and the greedy algorithms can be applied on the prelim-$l$ OS to find a good sub-optimal size-$l$ OS.

## 4. THE DP ALGORITHM

This section describes a dynamic programming (DP) algorithm, which, given an OS, determines the optimal size-$l$ OS in it. The OS is a tree, as discussed in Section 2. Every node $v$ of the OS tree is a tuple $t_i$, and carries a weight $w(v)$, which is the local importance $Im(OS, t_i)$ of the corresponding tuple $t_i$. Given the tree OS, our objective is to find a subtree $S_{opt}$, such that (i) $S_{opt}$ includes the root node $t^{DS}$ of OS, (ii) the tree has $l$ nodes, and (iii) its nodes have the maximum sum of weights for all trees that satisfy (i) and (ii). In the third condition, the sum of node weights corresponds to $Im(S_{opt})$, according to Equation 2. Since this is the maximum among all qualifying subtrees, $S_{opt}$ is the optimal size-$l$ OS.

Assume that the root $t^{DS}$ in $S_{opt}$ has a child $v$ and the subtree $S^v_{opt}$ rooted at $v$ has $i$ nodes. Then, $S^v_{opt}$ should be the optimal size-$i$ OS rooted at $v$. DP operates based on exactly this assertion; for each candidate node $v$ to be included in the optimal synopsis and for each number of nodes $i$ in the subtree of $v$ that can be included, we compute the corresponding optimal size-$i$ synopsis and the corresponding sum of weights. The optimal size-$i$ synopsis rooted at $v$ is computed recursively from precomputed size-$j$ synopses ($j < i$) rooted at $v$'s children; to find it, we should consider all synopses formed by $v$ and all size-$(i-1)$ combinations of its children and subtrees rooted at them.

Specifically, let $d(v)$ be the depth of a node $v$ in OS (the root $t^{DS}$ has depth 0). The subtree rooted at $d(v)$ can contribute at most $l-d(v)$ nodes to the optimal solution, because in every solution that includes $v$, the complete path from the root to $v$ must be included (due to the fact that $t^{DS}$ should be included and the solution must be connected). The construction of the DP algorithm is to compute for each node $v$ of OS $S_{v,i}$: the optimal size-$i$ OS for all $i \in [1, l - d(v)]$, in the subtree rooted at $v$. In addition to $S_{v,i}$ the algorithm should track $W(S_{v,i})$, the sum of weights of all nodes in $S_{v,i}$.

DP (Algorithm 1) proceeds in a bottom-up fashion; it starts from nodes in OS at depth $l - 1$; these nodes can only contribute themselves in the optimal solution (nodes at depth at least $l$ cannot participate in a size-$l$ OS). For each such node $v$, trivially $S_{v,1}=v$, $W(S_{v,1})=w(v)$. Now consider a node $v$ at depth $k<l-1$. Upon reaching $v$, for all children $u$ of $v$, quantities $S_{u,i}$ and $W(S_{u,i})$ have been computed for all $i \in [1, l - d(v) - 1]$. Let us now see how we can compute $S_{v,i}$ for each $i \in [1, l - d(v)]$. First, each $S_{v,i}$ should include $v$ itself. Then, we examine all possible combinations of $v$'s children and number of nodes to be selected from their subtrees, such that the total number of selected nodes is $i-1$. We do not have to check the subtrees of $v$'s children, since for each number of nodes $j$ to be selected from a subtree rooted at child $u$, we already have the optimal set $S_{u,j}$ and the corresponding sum of weights $W(S_{u,j})$. Note that when we reach the OS root $r = t^{DS}$, we only have to compute $S_{r,l}$: the optimal size-$l$ OS (i.e., there is no need to compute $S_{r,i}$ for $i \in [1, l-1]$).

**Algorithm 1** The Optimal Size-$l$ OS (DP) Algorithm

$DP(l, t^{DS}, G^{DS})$
1: **OS Generation**($t^{DS}, G^{DS}$)  ▷ generates the complete OS, annotates with local importance each node
2: **for each** node $v$ at depth $l-1$ **do** set $S_{v,1} = v$
3: **for each** depth $k = l - 2$ to 0 **do**
4:   **for each** node $v$ at depth $k$ **do**
5:     **for** $i=1$ to $l - d(v)$ **do**
6:       $S_{v,i} = \{v\} \cup$ the best combination of $v$'s children and nodes from them such that the total number of nodes is $i - 1$
7: **return** $S_{r,l}$

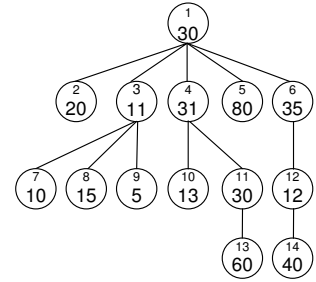

| Depth | Computed Sets |
|---|---|
| 3 | $S_{13,1}$=13, $S_{14,1}$=14 |
| 2 | $S_{7,1}$=7, $S_{8,1}$=8, $S_{9,1}$=9, $S_{10,1}$=10, $S_{11,1}$=11, $S_{11,2}$={11,13}, $S_{12,1}$=12, $S_{12,2}$={12,14} |
| 1 | $S_{2,1}$=2, $S_{3,1}$=3, $S_{3,2}$={3,8}, $S_{3,3}$={3,7,8}, $S_{4,1}$=4, $S_{4,2}$={4,11}, $S_{4,3}$={4,11,13}, $S_{5,1}$=5, $S_{6,1}$=6, $S_{6,2}$={6,12}, $S_{6,3}$={6,12,40} |
| 0 | $S_{1,4}$={1,4,5,6} |

**Figure 4: Example: Steps of DP**

As an example, consider the OS shown in Figure 4 (top) and assume that we want to compute the optimal size-4 OS from it. The table shows the steps of DP in computing the optimal sets $S_{v,i}$ in a bottom-up fashion, starting from nodes 13 and 14 which are at depth 3 (i.e. $l-1$). For example, to compute $S_{4,3}$={4,11,13}, we compare the two possible cases $S_{4,3} = \{4\} \cup S_{10,1} \cup S_{11,1}$ and $S_{4,3}=\{4\} \cup S_{11,2}$ since $S_{10,1} \cup S_{11,1}$ and $S_{11,2}$ are the only combinations sets from node 4's children that total to 2 nodes ($i-1$=2). $S_{10,1} \cup S_{11,1}$={10,11} with total weight 43 and $S_{11,2}$={11,13} with total weight 90. Thus, $S_{4,3} = \{4\} \cup S_{11,2}$={4,11,13}. Note that for nodes that do not have enough children, the number of sets that are computed could be smaller than those indicated in the pseudocode. For example, for node 2, we only have $S_{2,1}$; i.e. $S_{2,2}$ and $S_{2,3}$ do not exist although the node is at depth 1, because node 2 has no children. In addition, for the root node, DP only has to compute $S_{1,4}$, since we only care about the optimal size-$l$ summary (there are no nodes above the root that could use smaller summaries).

In terms of complexity, we need to compute for each node $v$ in the OS up to depth $l-1$ up to $l-d(v)$ sets. For each set we need to find the optimal combination of children and nodes from them to choose. This cost of choosing the best combination increases exponentially with $i$, which is O($l$). Thus, the overall cost of DP is $O(n^l)$ for an input OS of size $n$, as can be verified in our experiments. This is essentially the complexity of the problem as DP explores all possible summaries systematically and, in the general case, there is no way to prune the search space. For large values of $l$, DP becomes impractical and we resort to the greedy heuristics described in the next section. Finally, the following lemma proves the optimality of DP.

LEMMA 1. *Algorithm 1 computes the optimal size-l OS.*

PROOF. The optimal size-$l$ OS $S_{opt}$ includes the root $t^{DS}$ of the OS and a set of subtrees rooted as some of $t^{DS}$'s children. DP tests all possible combinations of children and numbers of nodes from the corresponding subtrees, therefore the combination that corresponds to $S_{opt}$ will be considered. For the specific combination, for each child $v$ and number of nodes $i$, the optimal subtree rooted at $v$ with $i$ nodes (i.e., $S_{v,i}$) has already been found during the bottom-up computation process of DP. Therefore, DP will select and output the optimal combination (which has the largest importance among all tested ones). □

## 5. GREEDY ALGORITHMS

Since the DP algorithm does not scale well, in this section, we investigate greedy heuristics that aim at producing a high-quality size-$l$ OS, not necessarily being the optimal. A property that the algorithms exploit is that the local importance of tuples in the OS (i.e. $Im(OS, t_i)$) usually decreases with the node depth from the root $t^{DS}$ of the OS. Recall that $Im(OS, t_i)$ is the product $Im(t_i) \cdot Af(t_i)$, where $Im(t_i)$ is the global importance of tuple $t_i$ and $Af(t_i)$ is the affinity of the relation that $t_i$ belongs to. $Af(t_i)$ monotonically decreases with the depth of the tuple since $Af(R_i)$ is a product of its parent's affinity and $Af(R_i) \leq 1$ (cf. Equation 1). On the other hand, the global importance for a particular tuple is to some extent unpredictable. Therefore, even though the local importance is not monotonically decreasing with the depth of the tuple on the OS tree, it has higher probability to decrease than to increase with depth. Hence, it is more probable that tuples higher on the OS to have greater local importance than lower tuples. Moreover, note that due to the non-monotonicity of OSs, existing top-$k$ techniques such as [6, 12, 17] cannot be applied.

### 5.1 Bottom-Up Pruning Size-$l$ Algorithm

This algorithm, given an initial OS (either a complete or a prelim-$l$ OS) iteratively prunes from the bottom of the tree the $n - l$ leaf nodes with the smallest $Im(OS, t_i)$, where $n$ is the number of nodes in the complete OS. The rationale is that since tuples need to be connected with the root and lower tuples on the tree are expected to have lower importance, we can start pruning from the bottom. A priority queue ($PQ$) organizes the current leaf nodes according to their local importance. Algorithm 2 shows a pseudocode of the algorithm and Figure 5 illustrates the steps.

More precisely, this algorithm firstly generates the initial OS (line 1; e.g. the complete OS using Algorithm 5). The OS Generation algorithm generates the initial size-$l$ OS and also the initial $PQ$ (initially holding all leaves of the given OS). Then, the algorithm iteratively prunes the leaves with the smallest $Im(OS, t_i)$. Whenever a new leaf is created (e.g., after pruning node 9 in Figure 5, node 3 becomes a leaf), it is added to $PQ$. The algorithm terminates when only $l$ nodes remain in the tree. The tree is then returned as the size-$l$ OS. In terms of time complexity, the algorithm performs O($n$) delete operations in constant time, each potentially followed by an update to the $PQ$. Since there are O($n$) elements in $PQ$, the cost of each update operation is O($\log n$). Thus, the overall cost of the algorithm is O($n \log n$). This is much lower than the complexity of the DP algorithm, which gives the optimal solution.

On the other hand, this method will not always return the optimal solution; e.g. the optimal size-5 OS should include nodes 1, 5, 6, 12 and 14 instead of 1, 5, 6, 11 and 13 (Fig 5(d)). In practice, it is very accurate (see our experimental results in Section 6.2), due to the aforementioned property of $Im(OS, t_i)$, which gives higher probability to nodes closer to the root to have a high local importance. Lemma 2 proves an optimality condition for this algorithm

**Algorithm 2** The Bottom-Up Pruning Size-$l$ Algorithm

*Bottom-Up Pruning Size-$l$* ($l, t^{DS}, G^{DS}$)
1: **OS Generation**($t^{DS}, G^{DS}$)  ▷ generates initial size-$l$ (i.e. complete or prelim-$l$) OS and initial $PQ$
2: **while** (|size-$l$ OS| > $l$) **do**
3: $\quad t_{tem}$=deQueue($PQ$)  ▷ the smallest value from $PQ$
4: $\quad$ **if** !(hasSiblings(size-$l$ OS, $t_{tem}$)) **then**
5: $\quad\quad$ enQueue($PQ$, parent(size-$l$ OS, $t_{tem}$))  ▷ check whether after pruning $t_{tem}$, its parent becomes a leaf node
6: $\quad$ prune $t_{tem}$ from size-$l$ OS
7: **return** size-$l$ OS

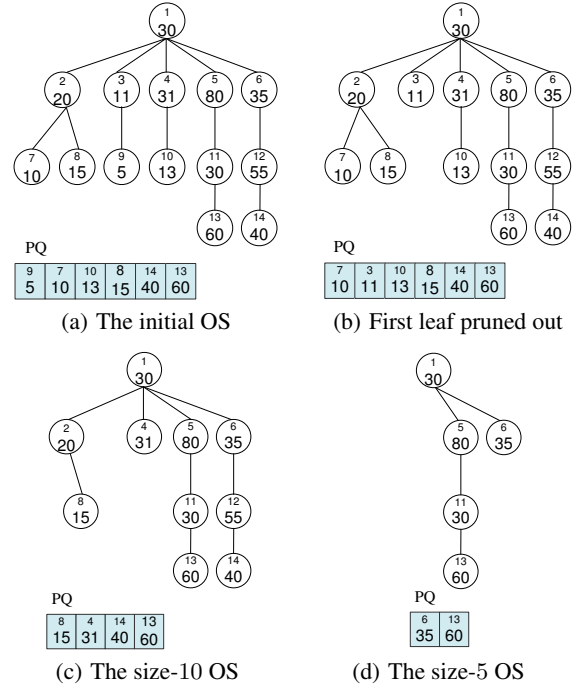

**Figure 5: The Bottom-Up Pruning Size-$l$ Algorithm: Size-$l$ OSs and their Corresponding $PQ$s (annotated with tuple ID and local importance)**

(Paper OSs in the DBLP database are an example of this condition; to be discussed in Section 6.2).

LEMMA 2. *When the nodes of an OS have monotonically decreasing local importance scores to their distance from the root (i.e. the score of each parent is not smaller than that of its children), then the Bottom-Up Pruning Size-$l$ Algorithm returns the optimal size-$l$ OS.*

PROOF. $PQ.top$ always holds the node with the current smallest score in the OS. This is because $PQ.top$ is by definition the smallest among leaf nodes, where leaf nodes always have smaller scores than their ancestors. Therefore, by removing the $n - l$ current smallest values (iteratively stored in $PQ.top$) from an OS, we can get the optimal size-$l$ OS. □

### 5.2 Update Top-Path-$l$ Algorithm

We now explore a second greedy heuristic. This algorithm iteratively selects the path $p_i$ of tuples with the largest *average importance per tuple* (denoted as $AI(p_i)$), adds $p_i$ to the size-$l$ OS and removes the nodes of $p_i$ from the OS and updates $AI(p_i)$ for the remaining paths accordingly. The rationale of selecting the path of tuples (instead of the tuple) with the current largest importance, is

**Algorithm 3** The Update Top-path-$l$ Algorithm

*Update Top-path-$l$*($l, t^{DS}, G^{DS}$)
1: **OS Generation**($t^{DS}, G^{DS}$) ▷ generates initial size-$l$ (i.e. complete or prelim-$l$) OS, annotates tuples with $AI(p_i)$)
2: **while** (|size-$l$ OS| < $l$) **do**
3:     $p_i$=path with max $AI(p_i)$
4:     add first $l-$|size-$l$ OS| nodes of $p_i$ to size-$l$ OS
5:     **if** (|size-$l$ OS| < $l$) **then**
6:       remove selected path $p_i$ from the tree
7:       **for each** child $v$ of nodes in $p_i$ **do**
8:          update $AI(p_i)$ for each node $t_j$ in the subtree rooted at $v$
9: **return** size-$l$ OS

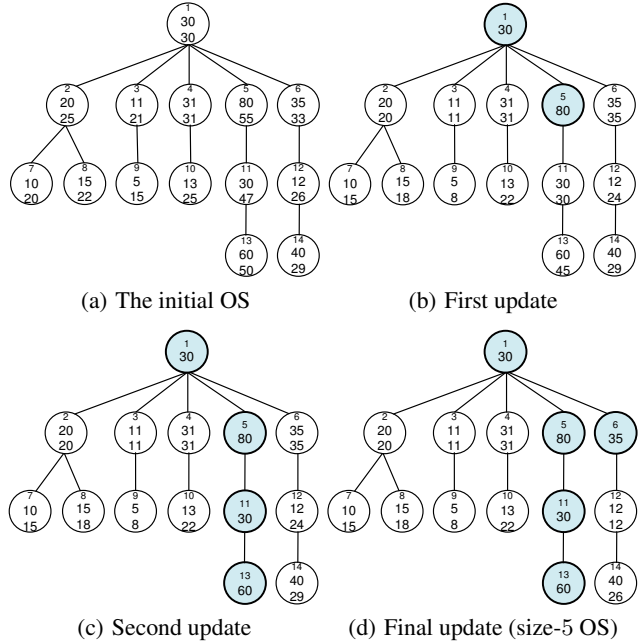

(a) The initial OS    (b) First update

(c) Second update    (d) Final update (size-5 OS)

**Figure 6: The Update Top-Path-$l$ Algorithm: The size-5 OS (annotated with tuple ID, local importance and $AI(p_i)$; selected nodes are shaded)**

that since all nodes need to be connected and monotonicity may not hold, we facilitate the selection of nodes of large importance even though their ancestors may have lower importance. Algorithm 3 is a pseudocode of the heuristic and Figure 6 illustrates an example.

More precisely, this algorithm (like the Bottom-Up Pruning Algorithm) firstly generates the complete (or alternatively the prelim-$l$) OS. During the OS generation, for each tuple $t_i$, we also calculate the importance per tuple $AI(p_i)$ for the corresponding path $p_i$ from the root to $t_i$. We then select the node with the largest $AI(p_i)$ and add the corresponding path to the size-$l$ OS. By removing the nodes of $p_i$ from the OS, the tree now becomes a forest; each child of a node in $p_i$ is the root of a tree. Accordingly, the $AI(p_i)$ for each node $t_i$ is updated again to disregard the removed nodes in the path selected at the previous step. The process of selecting the path with the highest $AI(p_i)$, adding it to the size-$l$ OS is repeated as long as less than $l$ nodes have been selected so far. If less than $|p_i|$ nodes are needed to complete the size-$l$ OS then only the top nodes of the path are added to the size-$l$ OS (because only these nodes are connected to the current size-$l$ OS).

Consider the example shown in Figure 6. Node 5 has $AI(p_i)$=55, because its path includes nodes 1 and 5 with average $Im(OS, t_i)$ being (30+80)/2=55. Assuming $l$=5, at the first loop, the algorithm selects nodes 1 and 5 with the largest $AI(p_i)$, i.e. 55. Then, the nodes along the path (nodes 1 and 5) are added to the size-5 OS. For the remaining nodes, $AI(p_i)$ is updated to disregard the removed nodes (see top-right tree in Figure 6). For example, the new $AI(p_i)$ for node 10 is 22, because its path now includes only nodes 4 and 10 with average $Im(OS, t_i)$ being 22. The next path to be selected is that ending at node 13, which adds two more nodes to the snippet. Finally, node 6 is added to complete the size-5 OS.

The complexity of the algorithm can be as high as $O(nl)$, where $n$ is the size of the complete OS, as at each step the algorithm may choose only one node which causes the update of $O(n)$ paths. The algorithm can be optimized if we precompute for each node $v$ of the tree the node $s(v)$ with the highest $AI(p_i)$ in the subtree rooted at $v$. Regardless of any change at any ancestor of $v$, $s(v)$ should remain the node with the highest $AI(p_i)$ in the subtree (because the change will affect all nodes in the subtree in the same way). Thus, only a small number of comparisons would be needed after each path selection to find the next path to be selected. Specifically, for each child $v$ of nodes in the currently selected path $p_i$, we need to update $AI(p_i)$ for $s(v)$ and then compare all $s(v)$'s to pick the one with the largest $AI(p_i)$. In terms of approximation quality, this algorithm not always returns the optimal solution; e.g. the size-3 OS will have nodes 1, 5 and 11 instead of 1, 5 and 6. However, empirically, this method gives better results than Bottom-Up Pruning.

### 5.3 Top-$l$ Prelim-$l$ OS Preprocessing

Instead of operating on the complete OS, which may be expensive to generate and search, we propose to work on a smaller OS, which hopefully includes a good size-$l$ OS. We denote such a preliminary partial OS as prelim-$l$ OS (with size $j$ where $l \leq j \leq |OS|$). On the prelim-$l$ OS, we can apply any of the proposed algorithms so far (of course, DP is not expected to return the optimal result, unless the prelim-$l$ OS is guaranteed to include it). The rationale of the prelim-$l$ OS is to avoid extraction and processing of tuples that are not promising to make it in the optimal size-$l$ OS. Algorithm 4 is a pseudocode for computing the prelim-$l$ OS, Table 1 summarizes symbols and definitions and Figure 7 illustrates an example.

Determining a prelim-$l$ OS that includes the optimal size-$l$ OS can be very expensive, therefore we propose a heuristic, which produces a prelim-$l$ OS that includes at least the $l$ nodes of the complete OS with the largest local importance (denoted as top-$l$ set). Figure 7(a) illustrates such a prelim-$l$ OS. Using *avoidance conditions* and simple *statistics* that summarize the range of local importance of every tuple in each relation (e.g. $\max(R_i)$) we can infer upper bounds for the local importance of tuples and thus safely predict whether a candidate path can potentially produce useful tuples.

DEFINITION 2. *Given an OS and an integer $l$, a top-$l$ prelim-$l$ OS (or simply prelim-$l$ OS) is a subset of the complete OS that includes the $l$ tuples of the OS with the largest local importance.*

We annotate each relation $R_i$ on the $G^{DS}$ graph with the statistics $\max(R_i)$ and $\text{mmax}(R_i)$ (see Figure 2). (Recall from Section 2.1 that we generate $G^{DS}$ graphs for every relation that may contain information about DSs.) $\max(R_i)$ is the maximum local importance of all tuples in $R_i$, which can be derived from the maximum global importance in $R_i$ (a global statistic that is computed/updated independently of the queries) and the affinity of $Af(R_i)$. $\text{mmax}(R_i)$ is the maximum local importance of all tuples that belong to $R_i$'s descendant relation nodes in $G^{DS}$ (i.e. the $\max_j\{\max(R_j)\}$; $j$ ranges over all such relations) or 0 if $R_i$ has no descendants (leaf node).

The algorithm for generating the prelim-$l$ OS is an extension of the complete OS generation algorithm (e.g. Algorithm 5). The extension incorporates pruning conditions in order to avoid adding to the prelim-$l$ OS fruitless tuples and their subtrees. More precisely,

**Table 1: Symbols and Definitions (Top-$l$ Prelim-$l$ OSs)**

| Symbols | Definition |
|---|---|
| top-$l$ | The $l$ nodes with the largest local importance in the OS |
| top-$l$ $PQ$ | An $l$-sized priority queue with the current largest local importance of extracted tuples |
| largest-$l$ | The tuple with the $l^{th}$ largest local importance retrieved so far (i.e. the smallest value of top-$l$ $PQ$) or 0 if \|top-$l$ $PQ$\|<$l$ |
| $li(t_i)$ | The local importance of tuple $t_i$(i.e. $Im(OS, t_i)$) |
| $R(t_i)$ | The relation on $G^{DS}$ that tuple $t_i$ belongs to |
| $R_i(t_j)$ | The subset of $R_i$ that joins with tuple $t_j$ |
| $max(R_i)$ | The maximum value of local importance of $R_i$ |
| $mmax(R_i)$ | The maximum value of $max(R_i)$ of all $R_i$'s descendants nodes on $G^{DS}$ or 0 if $R_i$ has no descendants (leaf node) |
| fruitless tuple | A tuple not in top-$l$ |
| fruitless $G^{DS}$ relation/ sub-tree | A $G^{DS}$ sub-tree starting from relation $R_i$ is considered *fruitless* for a given largest-$l$, if none tuples from $R_i$ and its descendants can be fruitful for the top-$l$(i.e. when largest-$l$≥$max(R_i)$ AND largest-$l$≥$mmax(R_i)$) |
| fruitful-$l$ relation | A relation $R_i$ is considered fruitful-$l$ for a given largest-$l$, if only up to $l$ nodes from the corresponding $R_i(t_j)$ can be fruitful for the top-$l$, (i.e. when largest-$l$≥$mmax(R_i)$) |

we traverse the $G^{DS}$ graph in a breath first order. Every extracted tuple is appended to the prelim-$l$ OS (lines 2 and 14) and to queue $Q$ (to facilitate the breadth first traversal of the $G^{DS}$; see lines 3 and 15). Let largest-$l$ be the tuple with the $l^{th}$ largest local importance retrieved so far. If the current tuple $t_i$ is greater than largest-$l$, $t_i$ is added to the $l$-sized priority queue top-$l$ $PQ$ as well (in order to update the top-$l$ set; lines 4 and 17). Largest-$l$ is set to the current smallest value of top-$l$ $PQ$ or to 0 if the top-$l$ $PQ$ does not contain $l$ values yet (lines 20-23). We traverse the $G^{DS}$ as follows. For each tuple de-queued from the queue $Q$ (line 6), we extract all its child nodes from each corresponding child relation (lines 7-12) and we employ the following avoidance conditions:

**Avoidance Condition 1 (Avoiding fruitless $G^{DS}$ sub-trees):** If the top-$l$ $PQ$ already contains $l$ tuples and largest-$l$ is greater than or equal to the local importance of all tuples of the current relation $R_i$ and all its descendants (i.e. largest-$l$≥$max(R_i)$ AND largest-$l$≥$mmax(R_i)$), then there is no need to traverse the sub-tree starting at $R_i$ (line 8). In such cases, we say that the sub-tree starting from $R_i$ is fruitless. For instance, consider the example of Figure 7; while retrieving tuple $y_8$, largest-$l$=0.37 and the current child relation $R_i$ is Conference with $max(R_i)$=0.22 and $mmax(R_i)$=0. Thus, we can safely infer that Conference has no fruitful tuples for the particular prelim-$l$ OS. This avoidance condition does not require any I/O operations as all information required can cheaply be obtained from the annotated $G^{DS}$.

**Avoidance Condition 2 (Limiting up to $l$ tuple extractions from fruitful-$l$ relations):** Assume that we are about to traverse $R_i$ in order to extract $R_i(t_j)$: the tuples in $R_i$ which join with the parent tuple $t_j$. We can limit the amount of tuples returned by this join up to $l$, if we can safely predict that none of their descendants (if any) can be fruitful for the top-$l$. We say a relation $R_i$ on the $G^{DS}$ is considered fruitful-$l$ for a given largest-$l$, if we can safely predict that only up to $l$ tuples from $R_i$ can be fruitful for the top-$l$ and none of their descendants (if any); this is the case when largest-$l$≥$mmax(R_i)$ but largest-$l$<$max(R_i)$. In other words, we can safely extract only up to $l$ tuples greater than the largest-$l$ from a fruitful-$l$ relation; i.e. there is no need to compute the complete join. For instance consider the example of Figure 7, where we are about to traverse the fruitful-$l$ relation PaperCitedBy (a leaf node on the $G^{DS}$, thus a fruitful-$l$ relation) in order to extract the joins with Paper tuple $p_2$. Then, we can extract from the database only up to $l$

**Algorithm 4** The Prelim-$l$ OS Generation Algorithm

*Prelim-$l$ OS Generation* $(l, t^{DS}, G^{DS})$

1: largest-$l$=0
2: add $t^{DS}$ as the root of the prelim-$l$
3: enQueue($Q, t^{DS}$)
4: enQueue(top-$l$ $PQ, t^{DS}$)
5: **while** !(IsEmptyQueue($Q$)) **do**
6:     $t_j$=deQueue($Q$)
7:     **for each** child relation $R_i$ of $R(t_j)$ in $G^{DS}$ **do**
8:         **if** !(largest-$l$≥$max(R_i)$ AND largest-$l$≥$mmax(R_i)$) **then**                                  ▷ Av. Cond. 1
9:             **if** (largest-$l$≥$mmax(R_i)$) **then**
10:                $R_i(t_j)$="SELECT * TOP $l$ FROM $R_i$ WHERE ($t_j$.ID=$R_i$.ID AND $R_i$.$li$ >largest-$l$)"      ▷ Av. Cond. 2.
                   $t_j$.ID and $R_i$.ID represent the keys that $t_j$ and $R_i$ join and
                   $R_i$.$li$ the local import. attribute of $R_i$
11:            **else**
12:                $R_i(t_j)$="SELECT * FROM $R_i$ WHERE ($t_j$.ID=$R_i$.ID)"
13:            **for each** tuple $t_i$ of $R_i(t_j)$ **do**
14:                add $t_i$ on prelim-$l$ as child of $t_j$
15:                enQueue($Q, t_i$)
16:                **if** ($li(t_i)$>largest-$l$) **then**
17:                    enQueue(top-$l$ $PQ, t_i$)
18:                    **if** (\|top-$l$ $PQ$\| >$l$) **then**
19:                        deQueue(top-$l$ $PQ$)
20:                    **if** (\|top-$l$ $PQ$\| <$l$) **then**
21:                        largest-$l$=0
22:                    **else**
23:                        largest-$l$=Smallest(top-$l$ $PQ$)

tuples with local importance greater than the largest-$l$ (which is 0, since \|top-$l$ $PQ$\|<$l$). Similarly, when traversing the fruitful-$l$ relation PaperCites with largest-$l$=0.12, we extract up to $l$ tuples larger than largest-$l$. Note that the Paper relation is not fruitful-$l$, since largest-$l$=0 and $mmax(R_{Paper})$=7.38 thus largest-$l$<$mmax(R_{Paper})$. As a consequence, we cannot apply this avoidance condition and hence we need to extract all tuples for Paper. Note, that this condition has no impact on M:1 relationships since the maximum cardinality of $R_i(t_j)$ is 1 anyway.

In terms of cost, in the worst case we need up to $n$ I/O accesses (if operating directly on the database), where $n$ is the amount of nodes in the complete OS, even if we extract only $j$ tuples (recall that Avoidance Condition 2 still requires an I/O access even when it returns no results). In practice, however, there can be significant savings if the top-$l$ tuples are found early and large subtrees of the complete OS are pruned. The prelim-$l$ OS created according to Definition 2 does not essentially contain the optimal size-$l$ OS, e.g. the prelim-5 OS of our example does not contain the $ca_{16}$ node which belongs to the optimal size-5 OS. In practice, we found that in most cases the prelim-$l$ OS did contain the optimal solution. This means that all size-$l$ OS computation algorithms may give the same results when applied either on the prelim-$l$ or complete OS. The following lemma proves that if monotonicity holds then the prelim-$l$ OS will certainly include the optimal size-$l$ OS.

LEMMA 3. *When the nodes of an OS have monotonically decreasing local importance scores to their distance from the root, then the prelim-l OS contains the optimal size-l OS.*

PROOF. When monotonicity holds, the optimal size-$l$ OS is the top-$l$ set (as shown by Lemma 2). Therefore, the prelim-$l$ OS produced by this algorithm that contains the top-$l$ set is optimal. □

Finally, we note that we have also investigated a variant of the prelim-$l$ OS, which includes the largest top-path-$l$ nodes (rather than the top-$l$), namely the $l$ tuples with the largest $AI(p_i)$. However, this approach did not result to better time or approximation quality so we do not further discuss it.

236

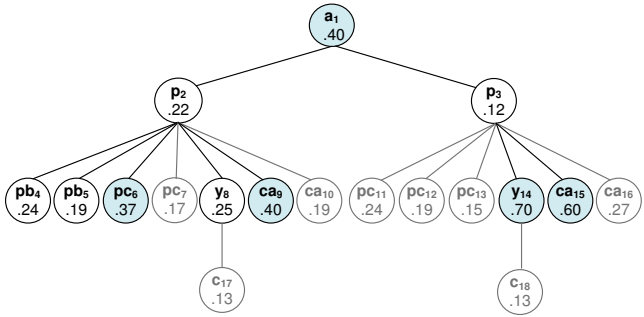

(a) The complete OS, the prelim-$l$ OS and the top-$l$ set. Nodes with low transparency are pruned tuples (e.g. $pc_7$, $ca_{10}$ etc.), shaded nodes are the top-$l$ set (e.g. $a_1$, $pc_6$ etc.) and the rest are the remaining tuples of the prelim-$l$ OS (e.g. $p_2$, $p_3$ etc.)

| $t_j$ | $R_i$ | $R_i(t_j)$ | $Q$ | top-$l$ $PQ$ | | | | | largest-$l$ |
|---|---|---|---|---|---|---|---|---|---|
| . | . | . | $a_1$ | $a_1$ .40 | | | | | 0 |
| $a_1$ | Paper | $p_2, p_3$ | $p_2, p_3$ | $a_1$ .40 | $p_2$ .22 | $p_3$ .12 | | | 0 |
| $p_2$ | Paper CitedBy | $pb_4, pb_5$ | $p_3, pb_4, pb_5$ | $a_1$ .40 | $pb_4$ .24 | $p_2$ .22 | $pb_5$ .19 | $p_3$ .12 | 0.12 |
| $p_2$ | Paper Cites | $pc_6$ (Av. Cond. 2) | $p_3, pb_4, pb_5$ $pc_6$ | $a_1$ .40 | $pc_6$ .37 | $pb_4$ .24 | $p_2$ .22 | $pb_5$ .19 | 0.19 |
| $p_2$ | Year | $y_8$ (Av. Cond. 2) | $p_3, pb_4, pb_5$ $pc_6, y_8$ | $a_1$ .40 | $pc_6$ .37 | $y_8$ .25 | $pb_4$ .24 | $p_2$ .22 | 0.22 |
| $p_2$ | Co-Author | $ca_9$ (Av. Cond. 2) | $p_3,\ldots, y_8$ $ca_9$ | $a_1$ .40 | $ca_9$ .40 | $pc_6$ .37 | $y_8$ .25 | $pb_4$ .24 | 0.24 |
| ... | ... | ... | ... | ... | ... | ... | ... | ... | ... |
| $y_8$ | Conference | . (Av. Cond. 1) | $ca_9, y_{14}$ $ca_{15}$ | $y_{14}$ .70 | $ca_{15}$ .60 | $a_1$ .40 | $ca_9$ .40 | $pc_6$ .37 | 0.37 |
| $ca_9$ | . (leaf) | . ($ca_9$ is leaf) | $y_{14}, ca_{15}$ | $y_{14}$ .70 | $ca_{15}$ .60 | $a_1$ .40 | $ca_9$ .40 | $pc_6$ .37 | 0.37 |
| $y_{14}$ | Conference | . (Av. Cond. 1) | $ca_{15}$ | $y_{14}$ .70 | $ca_{15}$ .60 | $a_1$ .40 | $ca_9$ .40 | $pc_6$ .37 | 0.37 |
| $ca_{15}$ | . (leaf) | . ($ca_{15}$ is leaf) | ∅ | $y_{14}$ .70 | $ca_{15}$ .60 | $a_1$ .40 | $ca_9$ .40 | $pc_6$ .37 | 0.37 |

(b) Values of $t_j$, $R_i$, $R_i(t_j)$, $Q$, top-$l$ $PQ$ and largest-$l$ during the prelim-$l$ OS generation

**Figure 7: The Prelim-$l$ OS Generation Algorithm ($l=5$, $t^{DS}=a_1$, and $G^{DS}=G^{Author}$)**

## 6. EXPERIMENTAL EVALUATION

In this section, we experimentally evaluate the proposed size-$l$ OS concept and algorithms. We evaluate our algorithms using both complete and prelim-$l$ OSs. First, the effectiveness of the proposed size-$l$ OSs is thoroughly investigated with the help of human evaluators. Then, the quality of the size-$l$ OSs produced by the greedy heuristics is compared to that of the corresponding optimal OSs. Finally, the efficiency of algorithms is comparatively investigated.

We used two databases: DBLP[2] and TPC-H[3] (we used scale factor 1 in generating the TPC-H dataset). The two databases have 2,959,511 and 8,661,245 tuples, occupying 319.4MB and 1.1GB on the disk, respectively.

We use ObjectRank (global) [3] and ValueRank [9] to calculate the global importance for the tuples of the DBLP and TPC-H databases respectively. For a more thorough evaluation, we investigate scores by various settings that have been studied in [3], namely, two $G^A$s: (1) the $G^{A1}$s (default) are presented in Figure 13 whereas (2) the $G^{A2}$ for the DBLP has common transfer rates (0.3) for all edges and for the TPC-H neglects values (i.e. becomes an ObjectRank $G^A$) and three values of $d$: $d_1=0.85$ (default), $d_2=0.10$ and $d_3=0.99$. We use Equation 1 to calculate affinity (alternatively an expert can define $G^{DS}$s and affinity manually, i.e. to select which relations to include in each $G^{DS}$ and their affinity). For the experiments, we used Java, MySQL, cold cache and a PC with an AMD Phenom 9650 2.3GHz (Quad-Core) processor and 4GB of memory.

[2] http://www.informatik.uni-trier.de/~ley/db/
[3] http://www.tpc.org/tpch/

### 6.1 Effectiveness

We used human evaluators to measure effectiveness. First, we familiarized them with the concepts of OSs in general and size-$l$ OSs in particular. Specifically, we explained that a good size-$l$ OS should be a stand-alone and meaningful synopsis of the most important information about the particular DS. Then, we provided them with OSs and asked them to size-$l$ them for $l$ = 5, 10, 15, 20, 25, 30. None of our evaluators were involved in this paper. Figure 8 measures the effectiveness of our approach as the average percentage of the tuples that exist both in the evaluators' size-$l$ OSs and the computed size-$l$ OS by our methods. This measure corresponds to *recall* and *precision* at the same time, as both the OSs compared have a common size.

**DBLP.** Since the DBLP database includes data about real people and their papers, we asked the DSs themselves (i.e. eleven authors listed in DBLP) to suggest their own Author and Paper size-$l$ OSs. The rationale of this evaluation is that the DSs themselves have best knowledge of their work and can therefore provide accurate summaries. Figures 8(a) and (b) plot the recall of the optimal size-$l$ OS for various ObjectRank settings. In general, ObjectRank scores produced with $G^{A1}$-$d_1$ and $G^{A1}$-$d_3$ are good options for Author and Paper size-$l$ OSs generation (as these settings produce similar ObjectRank scores) and always dominate on larger values of $l$. More precisely for $G^{A1}$-$d_1$, effectiveness ranges from 75% to 90% for $l$=10 to 30, and from 40% to 60% for $l$=5. These results are very encouraging. User evaluation also revealed that the inter-relational ranking properties (e.g. whether paper $p_1$ is more important than author $a_1$) affect crucially the quality of the size-$l$ OSs. For instance, on author OSs, evaluators first selected important Paper tuples to include in the size-$l$ OS and then additional tuples such as co-authors, year, conferences (these were usually included in summaries of larger sizes, i.e. $l{\geq}10$). The bias to select Papers (i.e., 1st-level neighbors) is favored by setting $G^{A1}$-$d_2$, although in overall this setting was not very effective; e.g., in Figure 8(a), this setting achieves 73.3% (in comparison to 60% of $G^{A1}$-$d_1$) for $l$=5.

The impact of approximated size-$l$ OSs produced by our greedy algorithms on effectiveness is very minor. For instance using scores produced by the default setting (i.e. $G^{A1}$ and $d_1$=0.85) on the Author $G^{DS}$, the Update Top-Path-$l$ algorithm generates summaries of the same effectiveness as the optimal, whereas Bottom-Up has very minor additional loss ranging from 2% to 10%. On the Paper $G^{DS}$, all approaches give the same effectiveness as they all return the optimal size-$l$ OSs. The use of prelim-$l$ OSs had no impact on effectiveness. As we show later, prelim-$l$ OSs have very minor impact on approximation quality which did not affect effectiveness.

**TPC-H.** We presented 16 random OSs to eight evaluators and asked them to size-$l$ them. The evaluators were professors and researchers from Manchester and Hong Kong Universities. In addition, for each OS and tuple, a set of descriptive details and statistics was also provided. For instance for a customer, the total number, size and value of orders and the corresponding minimum, median and maximum values of all customers were provided (e.g. similarly to the evaluation in [9]). The provision of such details gave a better knowledge of the database to the evaluators.

In summary, the $G^{A1}$ (for any $d$) is a safe option as it produces good size-$l$ OSs on both Customer and Supplier OSs (Figures 8(c) and 8(d)); e.g. effectiveness results for $G^{A1}$-$d_1$ range from 60% to 78%. On the other hand $G^{A2}$, which is the ObjectRank version of the $G^{A1}$, did not satisfy as much the evaluators on Supplier OSs.

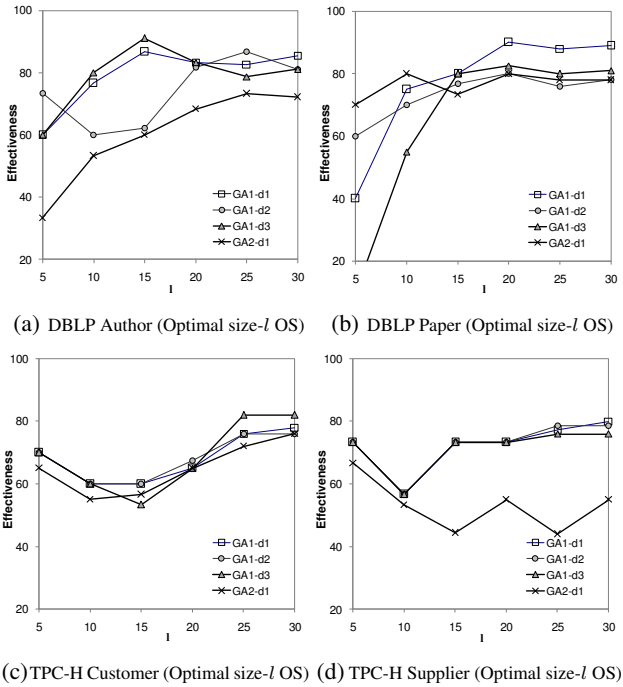

(a) DBLP Author (Optimal size-$l$ OS) (b) DBLP Paper (Optimal size-$l$ OS)

(c) TPC-H Customer (Optimal size-$l$ OS) (d) TPC-H Supplier (Optimal size-$l$ OS)

**Figure 8: Effectiveness (i.e. Recall=Precision)**

Interestingly, we observe that the effectiveness results for size 5 were very good on both OSs due to good inter-relational ranking.

**Comparative Evaluation.** We compared our results with Google Desktop (a text document search engine). We store each OS as an HTML file and then issue the corresponding query using Google Desktop in order to obtain its snippet. Google snippets contain a small amount of words from the beginning of the file, combining static text such as "Search for Christos Faloutsos in the DBLP Database" and the first few tuples (up to three) from the OS (note that the order of nodes in an OS is random). We make a less austere comparison by counting the selected tuples that belong to the corresponding size-5 OS proposed by our evaluators (since Google snippets contain only up to three tuples). As expected, in all cases Google snippets found zero and exceptionally one tuple from the corresponding size-5 OS. Detailed results are not shown due to space constraints.

## 6.2 Approximation Quality

We now compare the importance of the size-$l$ OSs produced by the greedy methods against the optimal ones. More precisely, the results of Figure 9 represent the approximation quality, namely the ratio of the achieved size-$l$ OS importance against the optimal importance. We present the average results for 10 random OSs per $G^{DS}$. The average size (i.e. the amount of tuples) of OSs is also indicated (denoted as Aver($|OS|$)).

Figures 9(a)-(e) show the approximation quality produced by the default settings (i.e. $G^{A1}$ and $d_1$=0.85). The results show that the Update Top-Path-$l$ is always better than the Bottom-Up Pruning algorithm. In general, the superiority of Update Top-Path-$l$ over Bottom-Up Pruning is up to 10% (excluding Paper OSs where all methods achieved 100%). The evaluation also reveals that top-$l$ prelim-$l$ OSs have very low approximation quality loss. They have no impact on the Bottom-Up algorithm and only up to 4% on the Update Top-Path-$l$ algorithm. Another observation is that the contents of the $G^{DS}$ and the values of the local importance scores also have a significant impact. For instance, for Paper OSs all methods achieved 100% quality. This is because the monotonicity property

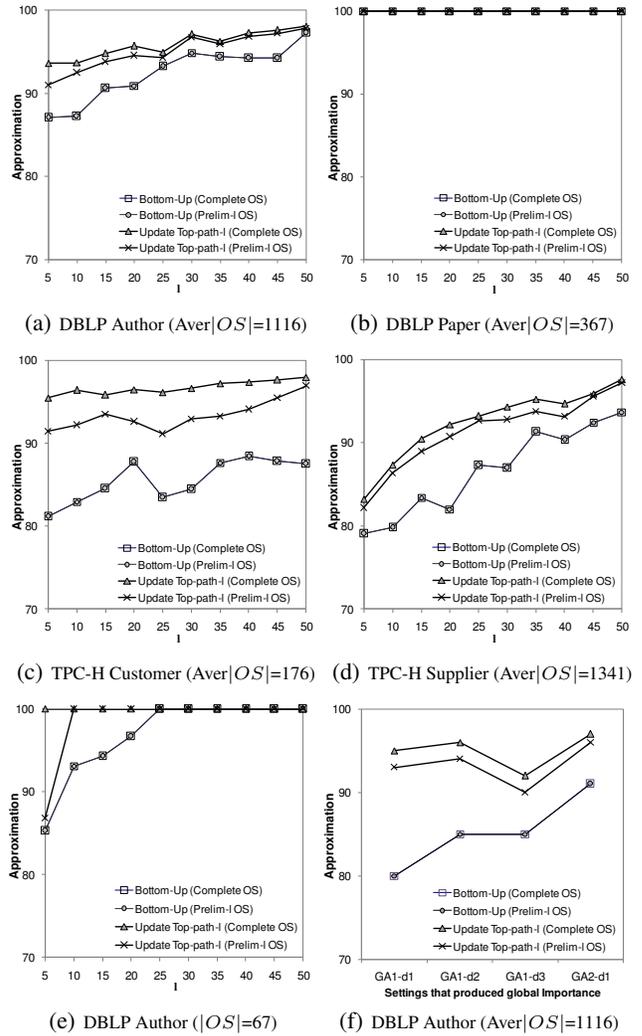

(a) DBLP Author (Aver$|OS|$=1116) (b) DBLP Paper (Aver$|OS|$=367)

(c) TPC-H Customer (Aver$|OS|$=176) (d) TPC-H Supplier (Aver$|OS|$=1341)

(e) DBLP Author ($|OS|$=67) (f) DBLP Author (Aver$|OS|$=1116)

**Figure 9: Approximation Quality**

holds (Lemma 2); the Paper $G^{DS}$ is Paper → (Author, PaperCitedBy, PaperCites, Year → (Conference)) and the local importance of Conferences is always smaller than those of the corresponding Years. In general, inter-relational and intra-relational ranking of tuples have an impact as well. For instance, Figure 9(f) summarizes the average approximation quality for Author OSs with global importance scores produced by the various settings (where inter and intra relational scores vary). The experimental results also reveal that the smaller the OS is in comparison to $l$ the more accurate our algorithms are. For example, the particular Author OS of Figure 9(e) with $|OS|$=67 yields 100% approximation quality from all algorithms, by $l$=25.

## 6.3 Efficiency

We compare the run-time performance of our algorithms in Figure 10. We used the same OSs as in Section 6.2 (i.e. the same 10 OSs per $G^{DS}$) and used the default settings for generating the global importance of tuples (alterative settings do not have any impact on the performance). Figures 10(a)-(e) show the costs of our algorithms for computing size-$l$ OSs from OSs of various sizes and different $l$ values, excluding the time required to generate the OS where each algorithm operates on. Figures 10(a)-(d) show the costs of OSs from various $G^{DS}$s, while Figure 10(e) shows scalability for Author OSs of different sizes and common $l$=10 (analogous re-

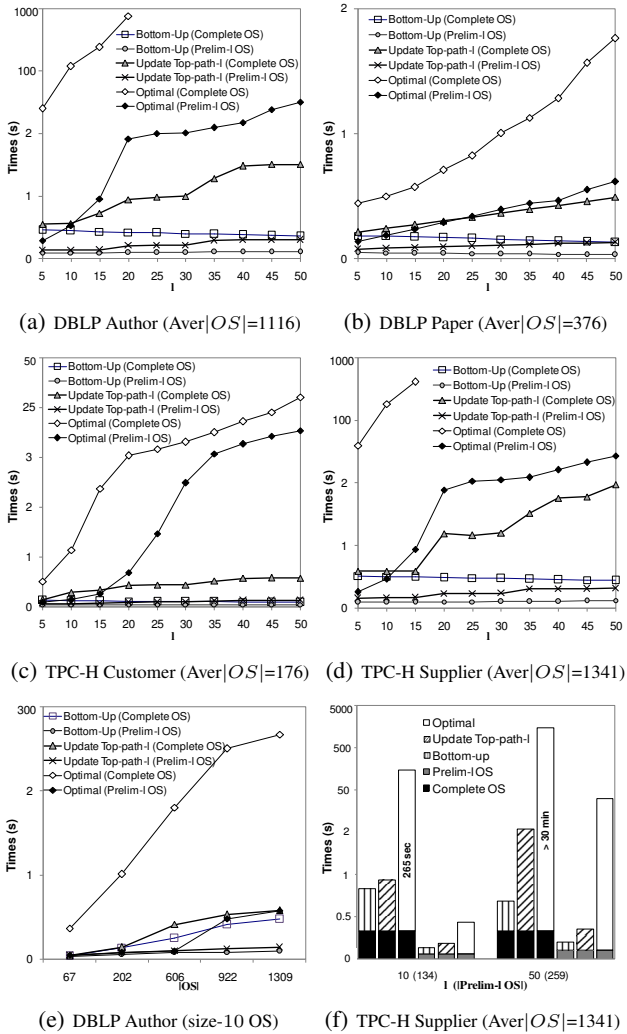

(a) DBLP Author (Aver$|OS|$=1116)
(b) DBLP Paper (Aver$|OS|$=376)
(c) TPC-H Customer (Aver$|OS|$=176)
(d) TPC-H Supplier (Aver$|OS|$=1341)
(e) DBLP Author (size-10 OS)
(f) TPC-H Supplier (Aver$|OS|$=1341)

**Figure 10: Efficiency**

sults were obtained from all $G^{DS}$s but we omit them due to space limitations). Note that the y-axes (time) in all graphs are split to two parts; one linear (bottom) and one exponential (top) in order to show how the expensive DP scales and at the same time keep the differences between the other methods visible.

As expected, the OS size and $l$ have affect significantly the cost (the bigger OS or $l$ is, the more time is required). The cost of DP becomes unbearable moderate to large OSs and values of $l$ (we had to stop the algorithms after 30 min. of running). Bottom-Up Pruning is consistently faster than Update Top-Path-$l$, as it requires fewer operations. An interesting observation is that Bottom-Up Pruning on the complete OS becomes faster as $l$ increases, because $n$-$l$ drops and fewer de-heaping operations are needed.

Figure 10(f) breaks down the cost to OS generation (bottom of the bar) and size-$l$ computation (top of the bar) for each method. We investigated two approaches for generating the OSs; the first employs an in-memory data-graph and the second computes the OS directly from the database. The OSs are generated much faster using the data graph; thus, we present only the data-graph based results in Figure 10(f). For example, to generate the Supplier OSs (that have the largest sizes among all tested OSs) only 0.2 sec. are required using the data-graph, compared to 12.9 sec. directly from the database. The DBLP and TPC-H data-graphs take only 17 sec. and 128 sec. to generate and occupy 150MB and 500MB, respectively. More precisely, our data-graph nodes correspond to the database tuples and edges to tuples relationships (through their primary and foreign keys). Note that the data-graph is only an index and does not contain actual data as nodes capture only keys and global importance. Figure 10(f) also shows the average sizes of the complete OSs (1,341) and the prelim-$l$ OSs (134 and 259 for $l = 10$ and $l = 50$, respectively). The prelim-$l$ OS generation is always faster than that of the complete OS; for instance the prelim-5 OS's size is approximately 10% of the size of the complete OS and its generation can be done up to 2.5 times faster (the savings are not proportional, because there can be many accesses to fruitless relations during the prelim-$l$ OS generation; i.e. Avoidance Condition 2 which still requires access to relations even when it returns no results); thus, prelim-$l$ OSs further reduce the time required by our algorithms. Bottom-Up Pruning becomes on average up to 5.7 times faster whereas the Update Top-Path-$l$ is up to 4.1 times. Note that the size of the database does not impact the OS generation time, because hash-maps are used to look-up the required nodes of an OS; we omit experimental results, due to space constraints.

**Discussion.** In summary, the DP algorithm is not practical on large OSs and $l$'s whereas our greedy algorithms are very fast and as we showed in Section 6.2, their results are of high approximation quality. Note that, in this paper, our main focus has been on optimizing the size-$l$ OS generation, not the OS generation cost (which we leave for further investigation as future work). In addition, the use of prelim-$l$ OSs is constantly a better choice over the complete OSs since they are always faster with a very minor quality loss. If we need to find the size-$l$ OS at a high speed, then the Bottom-Up Pruning is a good choice, since it is consistently the fastest method (e.g. using the prelim-50 for Supplier costs 0.12+0.12=0.24 sec.). If the OS had to be generated from the database, then the Update Top-Path-$l$ algorithm is preferable as it gives better quality and is insignificantly more expensive (e.g. 8.08+0.32 sec.).

## 7. CONCLUSION AND FUTURE WORK

We investigated the effective and efficient generation of size-$l$ OSs. First, we gave a formal definition of the size-$l$ OS, which targets the synoptic and stand-alone presentation of a large OS. We proposed a dynamic programming algorithm and two efficient greedy heuristics for producing size-$l$ OSs. In addition, we proposed a preprocessing strategy that avoids generating the complete OS before producing size-$l$ OSs. A systematic experimental evaluation conducted on the DBLP and TPC-H databases verifies the effectiveness, approximation quality and efficiency of our techniques.

A direction of future work concerns the further exploration of algorithms using hashing and reachability indexing techniques [18]. Another challenging problem is the combined size-$l$ and top-$k$ ranking of OSs. In addition, the selection of an appropriate value for $l$ is an interesting problem; a natural approach is to select $l$ based on the amount of attributes or words it will result, e.g. 20 attributes or 50 words. However, this approach results to the reformulation of the problem and we plan to investigate it. Finally, it is observed that, in the general case, optimal size-$l$ OSs for different $l$ could be very different. This prevents the incremental computation of a size-$l$ OS from the optimal size-$(l-1)$ OS, limiting pre-computation or caching approaches that could accelerate computation. In the future, we plan to experimentally analyze the space of optimal size-$l$ OSs and identify potential similarities among them that could assist their pre-computation and compression.

## 8. ACKNOWLEDGEMENTS

We would like to thank Vagelas Hristidis for providing us with his ObjectRank code and DBLP database and our evaluators for their generous help and comments (in particular, Dimitris Papadias, George Samaras and Chirstos Schizas). Finally, we thank the anonymous reviewers for their constructive comments.

## APPENDIX

**Algorithm 5** The OS Generation Algorithm

*OS Generation* ($t^{DS}$, $G^{DS}$)
1: add $t^{DS}$ as the root of the OS
2: enQueue($Q$, $t^{DS}$)  ▷ Queue $Q$ facilitates breath first traversal
3: **while** !(isEmptyQueue($Q$)) **do**
4:   $t_j$=deQueue($Q$)
5:   **for each** child relation $R_i$ of $R(t_j)$ in $G^{DS}$ **do**
6:     $R_i(t_j)$="SELECT * FROM $R_i$ WHERE ($t_j$.ID=$R_i$.ID)"
7:     **for each** tuple $t_i$ of $R_i(t_j)$ **do**
8:       add $t_i$ on OS as child of $t_j$
9:       enQueue($Q$, $t_i$)
10: **return** OS

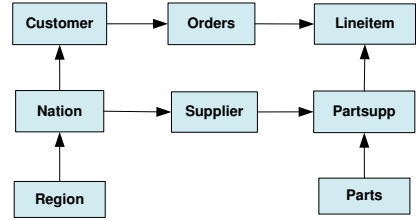

**Figure 11: The TPC-H Database Schema**

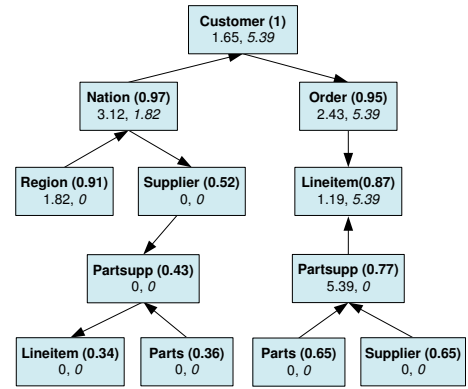

**Figure 12: The TPC-H Customers $G^{DS}$ (Annotated with (Affinity), max($R_i$) and mmax($R_i$))**

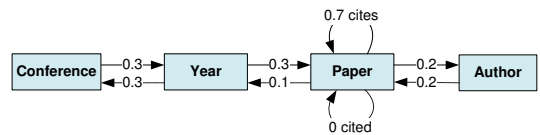

(a) The DBLP $G^A$

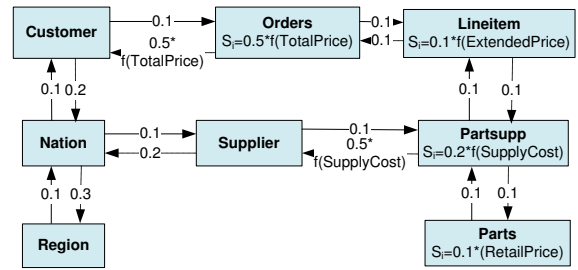

(b) The TPC-H $G^A$

**Figure 13: The $G^A$s for the DBLP and TPC-H Databases**